# Rapid, Quantitative Therapeutic Screening for Alzheimer's Enzymes Enabled by Optimal Signal Transduction with Transistors

Son T. Le[1,2], Michelle A. Morris[3,#], Antonio Cardone[4,5], Nicholas B. Guros[3,6,^],

Jeffery B. Klauda[6], Brent A. Sperling[7], Curt A. Richter[1], Harish C. Pant[8], and

Arvind Balijepalli[3,*]

[1]Alternative Computing Group, Nanoscale Device Characterization Division, National Institute of Standards and Technology, Gaithersburg, MD 20899, USA ; [2]Theiss Research, La Jolla, CA 92037; [3]Biophysics Group, Microsystems and Nanotechnology Division, National Institute of Standards and Technology, Gaithersburg, MD 20899, USA; [4]Information Systems Group, Software and Systems Division, National Institute of Standards and Technology, Gaithersburg, MD 20899, USA; [5]University of Maryland Institute for Advanced Computer Studies, University of Maryland, College Park, MD 20742, USA; [6]Department of Chemical and Biomolecular Engineering, University of Maryland, College Park, MD 20742, USA; [7]Chemical Process and Nuclear Measurements Group, Chemical Sciences Division, National Institute of Standards and Technology, Gaithersburg, MD 20899, USA; [8]National Institute of Neurological Disorders and Stroke, National Institutes of Health, Bethesda, MD 20892, USA
*e-mail: arvind.balijepalli@nist.gov
[#]Present address: Virginia Polytechnic Institute and State University, Blacksburg, VA 24061
[^]Present address: Astra Zeneca, 950 Wind River Lane, Gaithersburg, MD 20878

## Abstract

We show that simple, commercially sourced n-channel silicon field-effect transistors (nFETs) operating under closed loop control exhibit an ≈3-fold improvement in pH readout resolution to $(7.2\pm0.3)\times10^{-3}$ at a bandwidth of 10 Hz when compared with the open loop operation commonly employed by integrated ion-sensitive field-effect transistors (ISFETs). We leveraged the improved nFET performance to measure the change in solution pH arising from the activity of a pathological form of the kinase Cdk5, an enzyme implicated in Alzheimer's disease, and showed quantitative agreement with previous measurements. The improved pH resolution was realized while the devices were operated in a remote sensing configuration with the pH sensing element off-chip and connected electrically to the FET gate terminal. We compared these results with those measured by using a custom-built dual-gate 2D field-effect transistor (dg2DFET) fabricated with 2D semi-conducting $MoS_2$ channels and a moderate device gain, $\alpha=8$. Under identical solution conditions the pH resolution of the nFETs was only 2-fold worse than the dg2DFETs pH resolution of $(3.9\pm0.7)\times10^{-3}$. Finally, using the nFETs, we demonstrated the effectiveness of a custom polypeptide, p5, as a therapeutic agent in restoring the function of Cdk5. We expect that the straight-forward modifications to commercially sourced nFETs demonstrated here will lower the barrier to widespread adoption of these remote-gate devices and enable sensitive bioanalytical measurements for high throughput screening in drug discovery and precision medicine applications.

***Keywords:*** Field-effect transistor (FET), Complementary metal oxide semi-conductor (CMOS), 2D $MoS_2$, Biosensor, Super-Nernstian, pH, Enzyme activity, Enzyme therapeutics.

**Introduction**

Field-effect transistors (FETs) have a long history as sensitive and label-free bioanalytical tools.[1,2] Since that time FETs have been adapted for numerous applications that include measurements of protein-ligand interactions,[3] monitoring of ocean acidification,[4] low-cost DNA sequencing,[5] enzyme measurements,[6-8] and the detection of ionic action potentials in nerve and other neuronal systems.[9,10] In most cases, the FET-based sensors are fabricated by leveraging nanomanufacturing processes that underpin the silicon-based complementary metal-oxide semi-conductor (CMOS) industry. More recently, the emergence of 2D semi-conducting materials has resulted in new FET-based chemical sensors,[11,12] and novel device geometries such as dual-gate FETs, which provide ≈100-fold higher sensitivity than silicon devices, while simultaneously improving the signal-to-noise ratio (SNR) of the measurements.[8,13]

The sensitivity and resolution of pH measurements are important metrics of device function and are used, in particular, to validate the performance of new FET structures and designs.[8,14,15] Traditional ion sensitive field-effect transistors (ISFET) technology has been optimized to return a pH sensitivity that approaches the Nernst value of 59.5 mV when the solution pH changes by 1 at room temperature.[16] Efforts to improve the sensitivity of pH measurements have led to the exploration of dual-gate FETs for signal readout, which leverage the asymmetric capacitive coupling between the top- and back-gate with the device channel to amplify small pH signals measured remotely using commercial sensing elements. Recent measurements based on this approach have demonstrated the amplification of pH signals by 2-fold with silicon devices[15] and 37-fold (sensitivity of 2.25 V)[17] to 75-fold (sensitivity of 4.4 V)[8] higher than the Nernst sensitivity by using novel channel materials and gate structures. However, the improvement in sensitivity does not always result in better pH resolution for silicon devices because the noise level increases with the signal, as seen from recent theoretical[18] and experimental studies.[15] As a result, the pH resolution of silicon ISFETs range from $50\times10^{-3}$ for commercial ISFETs[4] to as low as $15\times10^{-3}$ for silicon nanowire devices.[19] On the other hand, dual-gate FETs fabricated with a 2D $MoS_2$ channel and with a room temperature ionic liquid top-gate dielectric have been demonstrated with a pH resolution as small as $92\times10^{-6}$ with a 10 Hz bandwidth.[8] In particular, these devices, which were operated in a low-noise regime of the atomically thin channel, exhibit a linear scaling in pH resolution improvement with device gain.

In order to drive adoption of high-resolution FET measurements within bioanalytical applications, we show how techniques developed for dual-gate FETs can be applied to commercially sourced silicon FETs. The techniques allow silicon devices to achieve pH readout resolution that exceeds most ISFET results and is on par with a solid-state version of recently published dual-gate 2D FETs (dg2DFETs).[8] Importantly, the improved performance of both nFETs and dg2DFETs was achieved in a remote sensing configuration where the pH sensitive surface is located off-chip and connected electrically to the FETs. The advantage of the remote sensing approach is the separation of signal transduction from the sensing surface. This allows reuse of the electronic components, minimizes parasitic noise sources, and enables measurements from myriad commercial and custom-built sensing elements. This setup also differs from most ISFET studies where the gate dielectric also serves as the pH sensing membrane.

We illustrate high-resolution pH measurements by comparing closed loop transduction of the signal from a commercial nFET to the open loop operation of the same device. The improved performance observed in the closed loop configuration establishes that this approach can allow the use of readily obtained commercially packaged transistors for laboratory grade bioanalytical measurements. Furthermore, the closed loop transduction approach can be applied to a wide range of sensor technologies that could be based on other stand-alone transistors such as junction FETs (JFETs) and bipolar junction transistors (BJTs) and even to integrated sensors such as ISFETs. It is expected that this readout approach will improve the performance of sensor systems based on any of these transistors. The technique is demonstrated through measurements of the activity and the effect of customized polypeptide therapeutics on the kinase Cdk5, which is implicated in Alzheimer's disease and numerous other debilitating neurological disorders.

**Experimental**

**n-Channel Silicon Field-Effect Transistors:** Commercially sourced (ALD110900PAL; Advanced Linear Devices, Sunnyvale, CA)[†] silicon n-channel field-effect transistors (nFETs) were soldered onto a printed circuit board (PCB) prior to measurements using a commercial probe station. Electrical characterization of the nFETs was performed using a semiconductor parameter analyzer (4155C; Agilent. Santa Clara, CA). Time-series measurements with the nFETs were performed similarly to the measurements done with dg2DFET as described below.

**2D Dual-Gate Transistor Fabrication:** A detailed description of the device fabrication process and monolayer characterization for dual-gate 2D field-effect transistors (dg2DFETs) was provided in our previous work.[8,13,20] Briefly, monolayer $MoS_2$ was first transferred onto an oxidized Si substrate ($SiO_2$ with a thickness of 70 nm) by using a gold-mediated exfoliation technique.[21] The thickness of the transferred material was confirmed with Raman spectroscopy.[13] Optical lithography was used to first pattern the source (*S*) and drain (*D*) contacts followed by electron-beam metal deposition (80 nm Au on 2 nm Ti) and lift-off in acetone. A second optical lithography step was then used to define and etch a 5 μm × 5 μm channel for each FET. The devices were then annealed under forming gas (5 % $H_2$, 95 % Ar) for 24 hours to minimize organic contamination and improve the contact resistance.[13] This was immediately followed by the atomic layer deposition (ALD) of a 20 nm top-gate (*TG*) $Al_2O_3$ dielectric. Finally, another optical lithography step was used to pattern the top-gate metal, followed by electron-beam metal deposition (100 nm Au on 10 nm Ti) and lift-off in acetone.

**Remote Biological Activity Measurements:** The enzyme and pH calibration measurements were performed with the nFETs and dg2DFETs were performed by connecting a pH sensor to the top-gate metal contact with a shielded coaxial cable. This remote sensing approach allows electronic components to be separated from the biological components and thereby reused. In the present work, a glass combination microelectrode (MI-4156; Microelectrodes, Bedford, NH) capable of measurement volumes as small as 50 µL was used as the pH sensor, although the techniques described here are compatible with other sensing and bioanalytical surfaces that can be electrically connected to the top-gate metal contact.

**Time-Series Field-Effect Transistor Measurements and PID Control:** Time-series measurements were performed by operating the nFETs and dg2DFETs under proportional-integral-derivative (PID) control (HF2LI; Zurich Instruments, Zurich, Switzerland). The channel current, $I_D$, was maintained at a constant value by continuously varying the back-gate voltage ($V_{BG}$) in the case of the dg2DFETs or by adding the controller output to the signal from the pH sensor ($V_{pH}$) for the nFETs in response to changes in the top-gate potential.

The PID control system was implemented by first amplifying the channel current, $I_D$, with a current preamplifier (DLPCA-200; FEMTO, Berlin, Germany) at a gain of either $10^6$ V/A (dg2DFET) or $10^3$ V/A (nFET). The output of the current preamplifier was then filtered through a 4-pole Bessel filter with a cutoff frequency of 5 kHz and sampled with a frequency of 25 kHz by using a 14-bit analog-to-digital converter (HF2LI; Zurich Instruments, Zurich, Switzerland).

The digital PID controller ($K_P$=553.5 mV, $K_I$=9.22×10³ s⁻¹ and $K_D$=10.4 µs) was operated with a bandwidth of 1 kHz to maintain the channel current set-point. Because most biological processes are slow and do not require high bandwidth measurements, the controller output was further filtered by using a low-pass filter with a cutoff frequency of 10 Hz prior to being recorded.

**Sensitivity and Resolution of pH Measurements:** The pH sensitivity and resolution were established as described in our previous work.[8] Briefly, a histogram from the raw $V_{PID}$ time-series data was computed for each measured pH. A sum of two Gaussian distribution functions was then fit to the histograms to obtain the peak positions and standard deviations of the reference potential and the measured pH signal. The difference in the peak positions between the pH and reference potentials ($\Delta V_{PID}$) was used to determine the pH sensitivity of the device. The measurement uncertainty ($\sigma_{PID}$) was then obtained by propagating the error when determining $\Delta V_{PID}$. For the nFETs, the pH resolution, $\Delta pH=(k \times \sigma_{PID})/V_{Nernst}$, is reported with expanded uncertainty ($k$=2), where $V_{Nernst}$ is the Nernst potential at room temperature. For the dg2DFETs, $\Delta pH=(k \times \sigma_{PID})/(\alpha \times V_{Nernst})$, where $\alpha$ is the device gain.

**Kinase Measurement Reagents:** The activity of Cdk5/p25 was estimated by measuring the phosphorylation of histone H1 as reported previously.[8] All measurements were performed with 18.5 nM of Cdk5/p25 (C0745; Sigma Aldrich, MO) in 1× kinase buffer to match physiological conditions using a volume of 50 µL. The substrate protein histone H1 (10223549001; Sigma Aldrich, MO) was suspended in deionized water at a stock concentration of 2 mg/mL and further diluted as described in the in the Results section. Substrate phosphorylation was initiated with a mixture of dithiothreitol (DTT) and adenosine triphosphate (ATP) with final concentrations of 250 µM and 5 mM respectively. The measurements were buffered using 5× kinase buffer, prepared by suspending 25 mM β-glycerol (G9422; Sigma Aldrich, MO), 50 mM MgCl₂ (5980;

Millipore, MA), 5 mM EGTA (E0396; Sigma Aldrich, MO), 2.4 mM EDTA (1002264786; Sigma Aldrich, MO), 1.25 mM MOPS (M1254; Sigma Aldrich, MO) in deionized water (DIW). The kinase buffer was then diluted in DIW to form the 1× kinase buffer used in the assays.

**Results and Discussion**

We electrically characterized nFET and dg2DFET devices to evaluate their performance and to determine the optimal operating conditions for biosensing applications. The devices were calibrated with standard pH buffer solutions to determine the gain, α, of the dg2DFET, and the noise performance, sensitivity, and resolution of both the nFETs and the dg2DFETs. Finally, both device types were used to measure the activity of the kinase Cdk5, an enzyme implicated in Alzheimer's disease, and to evaluate the effectiveness of a custom polypeptide, p5, as a therapeutic agent in modulating Cdk5 function.

**Silicon n-Channel FET Performance.** The nFETs (Fig. 1a; *right*) were first characterized electrically using the configuration shown in the schematic in Fig. 1a (*left*). The transfer characteristics of the device were determined by recording the drain current ($I_D$) as a function of the gate potential ($V_G$), while the drain voltage ($V_D$) was held constant. Fig. 1b shows the typical transfer characteristics ($I_D$—$V_G$) for the nFETs. The device exhibits up to ≈8 orders of magnitude change in $I_D$ when switching from an off-state to an on-state with a steep sub-threshold slope of ≈100 mV per decade at ≈300 K and gate leakage current ($I_G$) of ≈1 pA. Fig. 1b (*inset*) shows the transconductance ($g_m$), obtained by taking the numerical derivative of the transfer curve. The peak transconductance of the nFETs ($g_{m,peak}$) was found to be ≈78 μS at a voltage ($V_{gm,max}$) of 0.385 V.

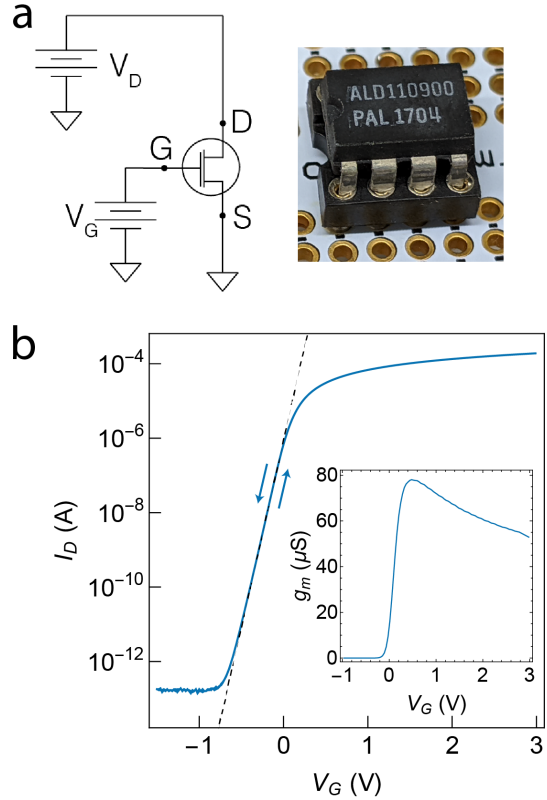

Figure 1: Electrical characterization of a commercially sourced silicon n-channel field-effect transistor (nFET) for biosensing applications. (a) *left*: Schematic of the electrical characterization setup of the nFET. A constant voltage, $V_D$, is applied to the drain contact while the source is grounded. The transfer characteristics of the device are obtained by sweeping the top-gate voltage, $V_G$, to electrostatically control the channel current ($I_D$). *right:* An image of the packaged nFET used in this work. (b) The transfer curve of the nFET is shown by measuring $I_D$ as a function of a sweep of $V_G$. The sub-threshold slope was found to be 100 mV per decade. (*inset*) The device transconductance as a function of $V_G$. The voltage at peak transconductance determines the point of maximum sensitivity and is used to optimally bias the device for biosensing.

**pH Sensitivity Using n-Channel Silicon Transistors.** The pH sensitivity of the nFETs was measured by using commercial standard pH buffer solutions over a range of 4 to 10. When operating the device in an open loop, shown schematically in Fig. 2a, we found the pH response to be linear when the device was operated about $g_{m,peak}$ as seen in Fig. 2b (*blue*). This behavior is expected when operating the device in the linear regime of the transfer curve (see Fig. 1). Under these conditions, we found the pH sensitivity, $\delta I_D/\delta pH \approx 6$ µA ($R^2=0.992$) when $V_D=0.1$ V, yielding a transimpedance gain of $9.9 \times 10^3$ V/A, assuming $V_{Nernst}$ of 59.5 mV at room temperature (300 K). The pH response was drastically different, as expected, when the device was operated about its threshold voltage ($V_T = 0$ V) as seen from Fig. 2b (*pink* and *inset*). In this case, the pH response was linear under acidic conditions (pH < 7), when the pH sensor returned a positive potential, thereby driving the FET into the inversion regime. On the other hand, under more basic conditions (pH > 7), the sensor potential was negative causing the nFET to operate in the sub-threshold regime where the current decreases exponentially at negative gate potential. The net result is a pH response that is highly non-linear over the measured pH range as has been observed by others in the literature.[22] Therefore, it may be advantageous to operate the nFET in the linear regime, particularly when operating over a wide pH range. Additionally, the larger $I_D$ in the linear regime over its value at $V_T$ results in lower relative noise and improved pH resolution as will be discussed in greater detail in later sections.

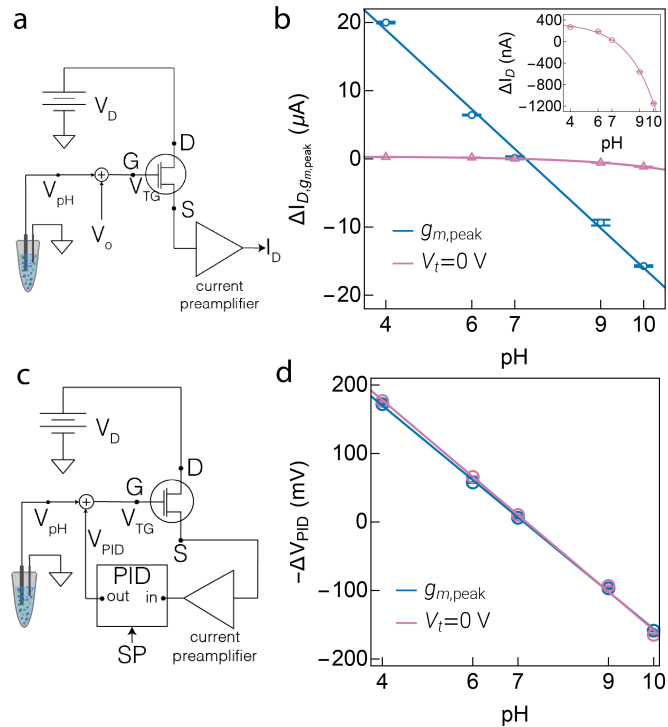

Figure 2: Calibration of pH measured with a commercially sourced n-channel silicon field-effect transistor (nFET). (a) The nFET was operated in open loop by directly measuring the changes in the drain current

($I_D$) due to changes in the gate potential ($V_{TG}$) applied to the top-gate contact (G). A constant offset potential ($V_o$) was summed with the potential ($V_{pH}$) generated by a glass pH microelectrode prior to being applied to the gate contact. (b) The change in $I_D$ as a function of measuring standard pH buffer solutions ranging from pH 4 to pH 10 with a glass pH microelectrode. Setting $V_o$ to operate the device at peak transconductance ($g_{m,peak}$) resulted in a linear response over the measured pH range (*blue*). Operating the device about the threshold voltage ($V_T$) of 0 V resulted in a highly non-linear pH response (*pink* and *inset*). (c) Constant current mode operation of the FET was performed by using a proportional-integral-derivative (PID) controller to monitor $I_D$ and continually adjust the PID voltage ($V_{PID}$). $V_{PID}$ was summed with $V_{pH}$ prior to being applied to the gate contact. (d) The change in $V_{PID}$ as a function of measuring standard pH buffer solutions ranging from pH 4 to pH 10 using a glass pH microelectrode. The PID set-point was set to a value of $I_D$ to allow device operation either at $g_{m,peak}$ (*blue*) or at the threshold voltage of 0 V (*pink*). Note that the error bars are smaller than the open circles that represent the data points. The error bars in (b) and (d) represent the expanded uncertainty ($k=2$) of the measurand.

The sensitivity and performance of the commercial nFET improved considerably when operated under closed loop PID control as shown schematically in Fig. 2c. In this configuration, the controller continually adds a control voltage ($V_{PID}$) to $V_{pH}$, thereby maintaining $I_D$ at a constant value. A key advantage of this approach is that because the device will always operate at the same point in its transfer curve (see Fig. 1b), its performance remains consistent across a wide range of measured pH values. This is clearly seen in Fig. 2d, where the device exhibits a linear pH response both when the PID controller was set up to hold the nFET at $g_{m,peak}$ (*blue*) and when the device was operating at its $V_T$ of 0 V (*pink*). Furthermore, the pH sensitivity in both cases, $dV_{PID}/dpH \approx 56$ mV ($R^2=0.997$), was obtained from a linear regression of the curves in Fig. 2d and approached the Nernst value of 59.5 mV at room temperature. Finally, the remote sensing configuration used here allows the PID output to be summed with the sensor signal, allowing the controller to operate at a higher bandwidth and allowing better noise suppression, thereby improving pH resolution.

**Dual-Gate 2D FET Performance.** We compared the performance of the nFETs with that of dg2DFETs that we fabricated from atomically thin $MoS_2$ films. An optical image of a representative dg2DFET device is shown in Fig. 3b. The dg2DFETs were electrically characterized following fabrication by using the setup shown schematically in Fig. 3a. The transfer characteristics of the device were measured by recording the drain current ($I_D$) as a function of the top-gate potential ($V_{TG}$) with the drain voltage ($V_D$) held constant (Fig. 3c). The measurements were repeated for different $V_{BG}$ to determine the signal amplification ($a$) due to the asymmetric capacitance of the top and back gates.[8] The devices exhibited a dynamic range of up to ≈5 orders of magnitude in $I_D$ and a subthreshold slope of ≈800 mV per decade, consistent with expected behavior for a 20 nm high-k $Al_2O_3$ gate dielectric.

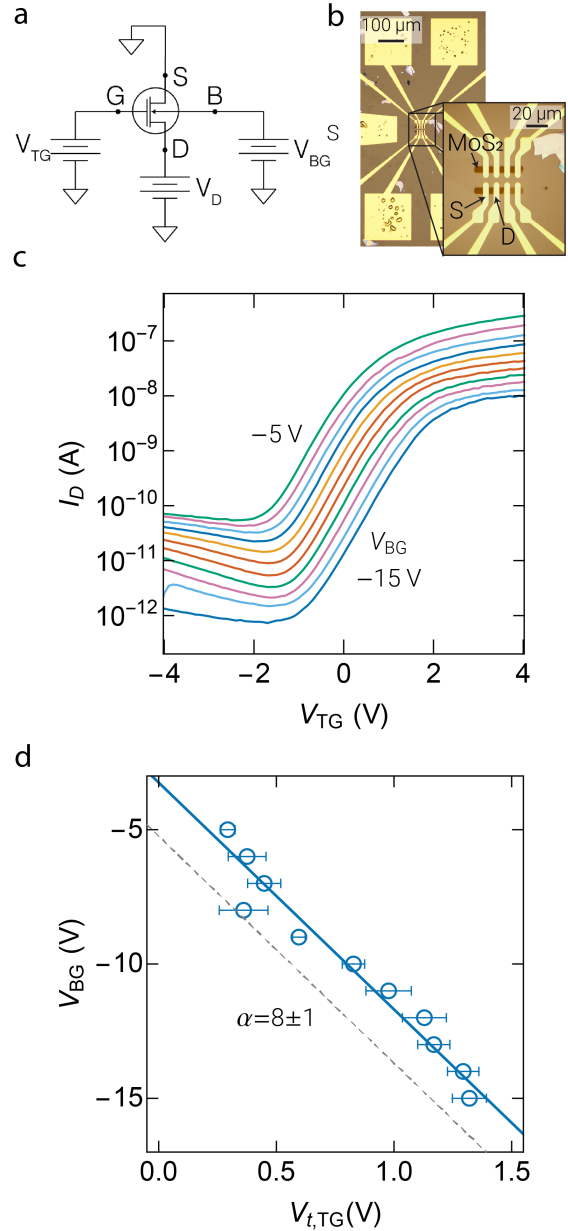

Figure 3: Electrical characterization of dual-gated monolayer MoS$_2$ field-effect transistors (FETs) for biological sensing applications. (a) Measurement schematic for characterizing a dual-gated 2D-FET for remote biosensing. The MoS$_2$ 2D semi-conducting channel spans the source (S) and drain (D) contacts. While the source contact is grounded, a constant potential ($V_D$) is applied to the drain contact driving a current across the 2D channel. The channel conduction is electrostatically controlled by a voltage applied to the silicon substrate, which forms the global back-gate (B) or to the metal top-gate (G). (b) Top view optical image of an array of 2D MoS$_2$ FETs. (c) Transfer characteristics of a dual-gated 2D FET showing drain current ($I_D$) as function of the top-gate voltage ($V_{TG}$) while stepping back-gate voltage ($V_{BG}$). (d) The change in $V_{BG}$ as function of top-gate threshold voltage ($V_{t,TG}$) is shown. A linear regression to the data (n=5) is used to determine the signal amplification ($\alpha$) of $V_{TG}$ at the back-gate. The error bars report the standard error defined as the standard deviation of the population mean.

For each curve in Fig. 3c, the top-gate threshold voltage ($V_{t,LG}$) was determined from a linear extrapolation of $I_D(V_{TG})$ at the voltage corresponding to peak transconductance to $I_D=0$.[23] Fig. 3d plots the back-gate voltage ($V_{BG}$) against the top-gate threshold voltage ($V_{t,TG}$). This allowed the determination of the device gain from the expression $\alpha=dV_{BG}/dV_{t,TG}$.[8] The value of $\alpha$ for devices measured as part of this work was then determined numerically from a linear regression to the data in Fig. 3d, resulting in $\alpha=8\pm1$, or ≈4 times larger than state-of-the-art dual-gate silicon devices.[15] The measured value of $\alpha$ is in good agreement with theoretical predictions for devices with a 20 nm $Al_2O_3$ top-gate dielectric and 70 nm $SiO_2$ bottom oxide.

It is important to note that the dg2DFETs presented here do not operate in the quantum capacitance limited regime as we showed previously with dual-gated ionic liquid devices.[8] This is due to the fact that, in the inversion regime where the dg2DFETs operate, the top-and back-gate capacitances ($C_{TG}\approx0.4$ µF/cm² and $C_{BG}\approx0.05$ µF/cm²) are more than an order of magnitude smaller than the quantum capacitance ($C_Q\approx4$ µF/cm²) of the 2D channel.[24,25] This allows us to ignore the effects of $C_Q$, giving rise to the simplified expression for the device gain, $\alpha=dV_{BG}/dV_{t,LG}=C_{TG}/C_{BG}$ (see Ref. 8 for a detailed derivation).

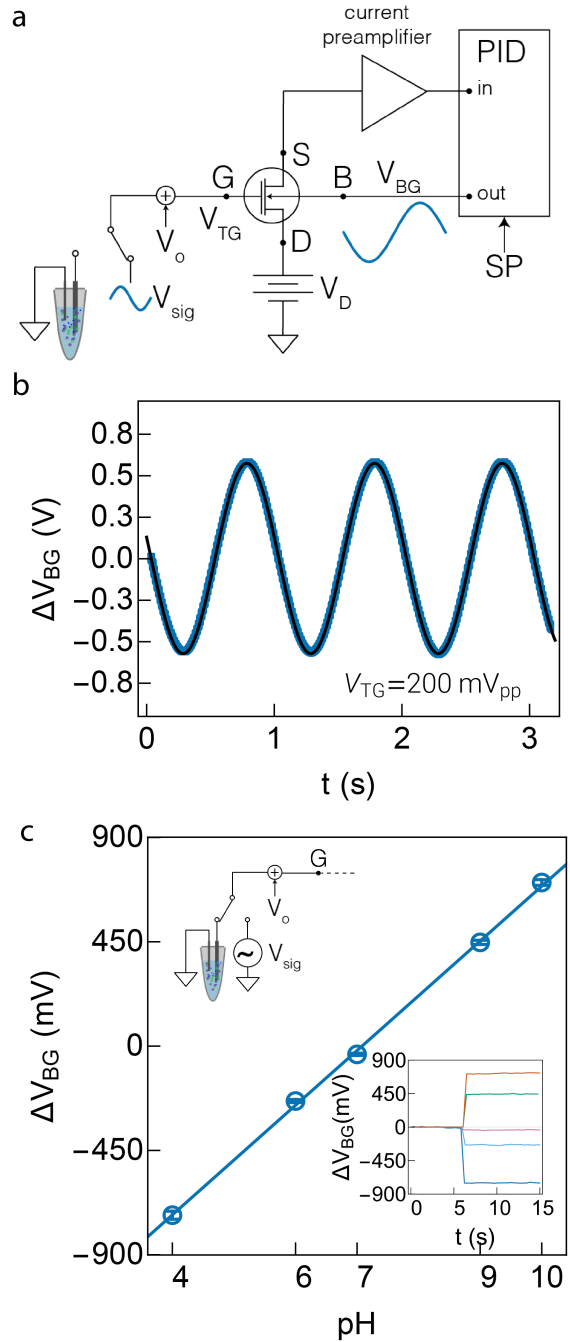

Figure 4: Electrical calibration and pH sensitivity measurements of dual-gate 2D field-effect transistors (dg2DFETs) when operated in a constant current mode. (a) Schematic representation of constant current mode operation of dg2DFETs.[8] A proportional-integral- derivative (PID) controller was used to maintain the channel current ($I_D$) at a constant value. Control of $I_D$ was achieved by continually adjusting the back-gate voltage ($V_{BG}$) in response to changes in the top-gate potential ($V_{TG}$) applied either using a waveform generator or from the output of a pH sensor ($V_{sig}$). A DC offset voltage ($V_o$) was summed with $V_{sig}$ to determine the optimal operation region of the dg2DFET. (b) The time-variant response of $V_{BG}$ under PID control is shown when the top-gate is biased with a 1 Hz AC sine wave signal with a peak-to-peak amplitude of 200 mV. (c) Response of $V_{BG}$ when measuring standard buffer solutions from pH 4 to 10.

The error bars represent the expanded uncertainty ($k=2$) of the measurement. Note that the error bars are smaller than the open circles that represent the data points. (*inset*) Time-series data, relative to a reference potential value, show the response of $V_{BG}$ when measuring standard buffer solutions from pH 4 to 10.

**pH Sensitivity of Dual-Gate 2D FETs.** The dg2DFETs were operated in a constant current mode as described in the *Experimental* section and shown schematically in Fig. 4a. A PID controller was used to maintain $I_D$ at a preset value by continuously varying $V_{BG}$ in response to changes in $V_{TG}$. The controller current set-point (50 nA) and a DC offset voltage applied to the top gate ($V_o$=+0.5 V) were optimized to operate the device in the linear and low-noise region of the transconductance curves in Fig. 3c. The performance of the device was then validated by applying a sine wave at a frequency of 1 Hz and peak-to-peak amplitude of 200 mV to the top-gate and measuring the response of $V_{BG}$ (Fig. 4b) as regulated by the PID controller. The device gain, $\alpha$, was then obtained from the ratio of the amplitudes of $V_{BG}$ to $V_{TG}$ to yield $\alpha$=5.8±0.1, consistent with the data in Fig. 3d.

The pH sensitivity of the dg2DFETs was measured by remotely connecting a pH sensor to the top-gate metal contact with a shielded cable. A switch was used to alternatively either ground the top-gate or connect it to the pH sensor. Time-series measurements of the system response, under PID control, to commercial standard buffer solutions from pH 4 to pH 10 are shown in Fig. 4c (*inset*). The time-series data were analyzed as described in the *Experimental* section to yield the pH response curve in Fig. 4c with a sensitivity, $dV_{BG}/dpH$= 236.3 mV ($R^2$=0.998).

**Comparison of pH Resolution of n-Channel Silicon and Dual-Gate MoS$_2$ Transistors.** The pH resolution ($\Delta pH$) of nFETs and dg2DFETs at a bandwidth of 10 Hz were determined when measuring phosphate buffered saline (PBS) and are summarized in Table 1. Following a procedure we developed previously,[8] a switch was used to alternatively either connect the gate terminal to the PBS solution or to ground. This method allowed the measurements of time-series of nFETs operated in an open loop as seen in Fig. 5a (*inset*). A histogram of the time-series was then used to determine the mean value of $I_D$ for each measured pH solution and the expanded uncertainty ($k$=2) as described in the *Experimental* section. As seen from Fig. 5a, a linear regression of the measured pH data yielded a sensitivity, $dI_D/dpH \approx 4$ µA ($R^2$=0.974) when $V_D$=0.1 V, similar to the value obtained from the data in Fig. 2b. By inverting the curve and propagating the uncertainty in $I_D$, we determined $\Delta pH$ with an expanded uncertainty ($k$=2) to be $(22 \pm 2) \times 10^{-3}$ at a bandwidth of 10 Hz in the open loop configuration.

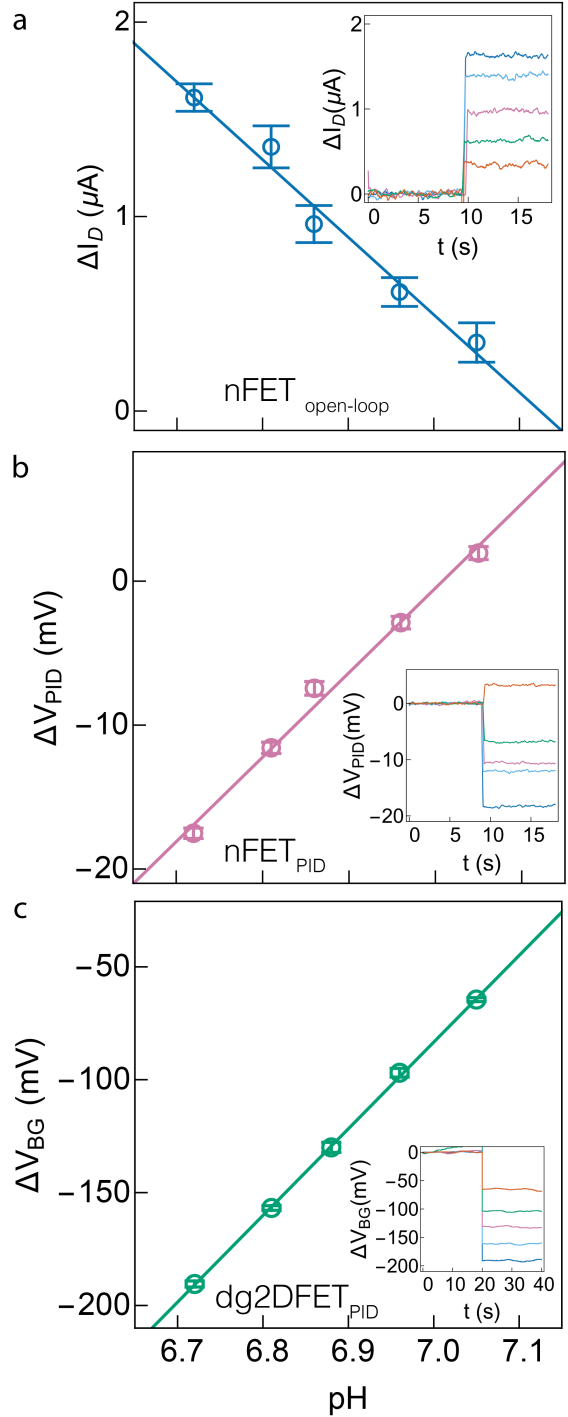

Figure 5: A comparison of pH sensitivity and resolution between a commercially sourced n-channel silicon field-effect transistor (nFET) operating open loop, operating under closed loop proportional-integral-derivative (PID) control, and a 2D dual-gated MoS$_2$ transistor (dg2DFET) operating under closed loop PID control. All measurements were performed with a pH sensitive glass microelectrode. (a) The change in the nFET channel current ($\Delta I_D$) when operating the device in open loop, as a function of phosphate buffered saline (PBS) solutions adjusted to different pH values. The pH sensitivity in this case was $dI_D/dpH \approx 4$ μA ($R^2$=0.974) when the drain voltage, $V_D$, was 0.1 V (b) The change in the PID control voltage ($\Delta V_{PID}$) as a function of solution pH when operating the nFET devices with PID control. In this case, the pH sensitivity was $dV_{PID}/dpH$=58.7 mV ($R^2$=0.988). (c) The change in the back-gate voltage, $\Delta V_{BG}$, as a function of solution pH when operating dg2DFETs under PID control. The pH sensitivity of the measurements, $dV_{BG}/dpH$ was found to be 384 mV ($R^2$=0.999) (*insets*) Underlying time-series data from nFETs and dg2DFETs that were analyzed to obtain the plots in panels (a), (b) and (c). The error bars in all cases represent the expanded uncertainty ($k$=2) of the measured quantity shown on the *y*-axis of the plot. Note that in (b) and (c) the error bars are smaller than the open circles that represent the data points.

The measurements were repeated when operating the nFET under PID control. The time-series of the PID controller output (Fig. 5b, *inset*) were analyzed identically to the open loop data to yield a pH sensitivity, $dV_{BG}/dpH$=58.7 mV ($R^2$=0.988) as seen from Fig. 5b, consistent with the expected value of $V_{Nernst}$ at room temperature. The error bars at each measured pH value, which represent the expanded uncertainty ($k$=2) in the PID output voltage, are a direct measure of $\Delta pH$. We found that, on average, $\Delta pH$=(7.2±0.3)×10$^{-3}$ at a bandwidth of 10 Hz or an ≈3-fold improved over nFETs operating in open loop.

Table 1: The smallest detectable pH change ($\Delta pH$) of phosphate buffered saline solution measured with a bandwidth of 10 Hz. The values were determined from the error bars reported in Fig. 5.

| Nominal pH | 6.72 | 6.81 | 6.88 | 6.96 | 7.05 |
| --- | --- | --- | --- | --- | --- |
| $\Delta pH$ nFET$_{open-loop}$ | 0.018 | 0.027 | 0.024 | 0.019 | 0.025 |
| $\Delta pH$ nFET$_{PID}$ | 0.006 | 0.007 | 0.008 | 0.007 | 0.008 |
| $\Delta pH$ dg2DFET$_{PID}$ | 0.003 | 0.003 | 0.006 | 0.005 | 0.002 |

Both modes of nFET operation described above were compared with pH measurements performed by using dg2DFETs in a closed loop configuration. Fig. 5c (*inset*) shows time-series measurements of the pH sensitivity of PBS buffers acquired by dg2DFETs when they were operated using the PID control scheme in Fig. 4a. An analysis of the pH time-series (see *Experimental* section for details) yielded a sensitivity, $dV_{BG}/dpH$=384 mV ($R^2$=0.999) as seen from Fig. 5c, which represents an ≈6.5-fold amplification of $V_{Nernst}$ at room temperature. The error bars in the figure estimate the expanded uncertainty ($k$=2) of the $\Delta V_{BG}$ at each pH value. This in turn allowed the estimation of the pH resolution from the expression $\Delta pH=\Delta V_{BG}/(\alpha \times V_{Nernst})$=(3.9±0.7)×10$^{-3}$ at a bandwidth of 10 Hz, or ≈2-fold better than the nFET devices operating under PID control.

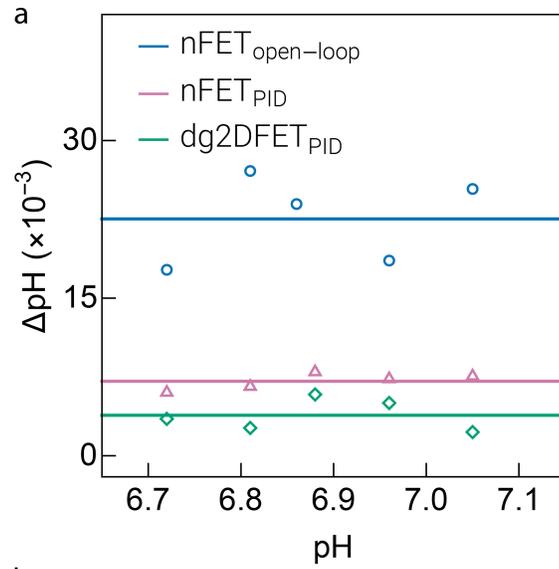

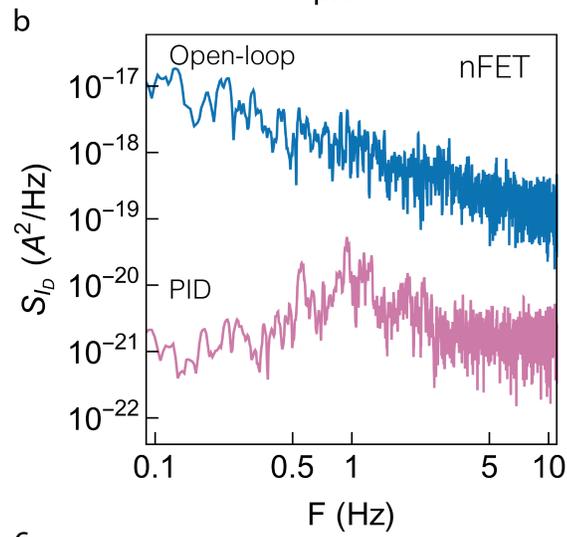

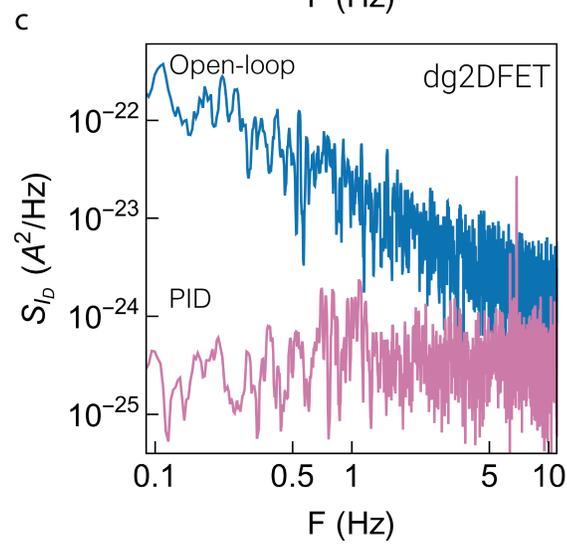

Figure 6: (a) A comparison of the pH resolution (*ΔpH*) as a function of pH when operating the n-channel silicon field-effect transistor (nFET) in open loop (*blue*), the nFET under PID control (*pink*) and the 2D dual-gated $MoS_2$ transistor (dg2DFET) under PID control (*green*). (b) Power spectral density (PSD) of the channel current, $I_D$, of nFETs under open loop operation (*blue*) and under PID control (*pink*). (c) Power spectral density (PSD) of the channel current, $I_D$, of dg2DFETs s under open loop operation (*blue*) and under PID control (*pink*).

As seen from Fig. 6a (*blue* and *pink*) and Table 1, the pH resolution of nFETs can be substantially improved by operating them under PID control. Under this mode of operation, the low intrinsic noise and the high channel currents of the nFETs allow their performance to approach that of custom-built dg2DFETs (Fig. 6a, *green*). To better understand the improvement in nFET performance under PID control, we measured the power spectral density (PSD) of the channel current noise, $S_{I_D}$, as seen in Fig. 6b. Under open loop operation (Fig. 6b, *blue*), the devices exhibit 1/f noise scaling as observed by others in the literature.[15,26] The root mean squared (RMS) noise in the channel current ($\delta I_D$) was then determined from the expression $\sqrt{\int_{BW} S_{I_D}\, df}$ to be 2.5 nA in the open loop case for a bandwidth of 10 Hz. The PID controller greatly suppresses 1/f noise as seen from Fig. 6b (*pink*). When operating under PID control, we found $\delta I_D$ to be 0.18 nA, or an order of magnitude lower for a bandwidth of 10 Hz. This reduction directly results in the improved *ΔpH* seen in Fig. 6a (*blue* and *pink*).

An improvement in the channel current noise is also observed for the dg2DFETs when operating under PID control, in comparison with the open loop case, as seen from Fig. 6c. At a bandwidth of 10 Hz, $\delta I_D$ decreased from 7.0 pA for open loop operation to 2.1 pA when operating under PID control. However, in contrast to the nFETs, the dual-gate devices have an intrinsic gain of ≈6.5, which improves their overall performance. In order to directly compare the nFETs with dg2DFETs when both devices were operating under PID control, we determined the relative noise for each case as $\delta I_D/I_D$, where the channel current, $I_D$, is also the current set-point for the PID controller. For the nFETs this value was 40.5 µA to allow operation at $g_{m,peak}$, and 50 nA for the dg2DFETs. This results in $\delta I_D/I_D = 4.3 \times 10^{-6}$ for the nFETs, and $\delta I_D/I_D = 4.2 \times 10^{-5}$ for the dg2DFETs. The order of magnitude lower relative noise in the nFETs is offset by the gain, $\alpha$, resulting from the dual-gated structure of the dg2DFETs and is consistent with the measured improvement in $\Delta pH$ seen in Fig. 6a (nFET$_{PID}$; *pink* and dg2DFET$_{PID}$; *green*). Furthermore, as noted earlier, the remote sensing configuration and the use of commercial devices for signal transduction implies that the performance of this system can be continually improved by substituting the existing commercial FETs with other lower noise devices, which in turn can improve $\Delta pH$.

While the pH resolution of the nFETs can approach that of the dg2DFETs with a moderate internal gain, their performance is expected to fall short of high gain devices such as ionic liquid gated dual-gate FETs.[8] The highly asymmetric gate geometries of the ionic liquid gated FETs allows the realization of $\alpha > 150$ while operating in a low-noise regime similar to the dg2DFETs. The combination of high gain and low noise allows those devices to resolve pH values as small as $92 \times 10^{-6}$, which is an order of magnitude below the resolution attainable by the dg2DFETs (Fig. 6a). While this improved resolution is critical in certain bioanalytical applications, we show below that the improvements in the operation modes of nFETs demonstrated here can be leveraged to measure both enzymatic activity and the effect of therapeutics on enzyme function at physiological concentrations.

**Cdk5-p25 Pathological Activity and Neurodegeneration.** The cyclin-dependent kinase 5 (Cdk5) is of fundamental importance for neuronal development, memory and pain signaling.[27-30] Its physiological activators, the proteins p35 and p39, trigger the Cdk5-mediated phosphorylation of neuronal proteins and organelles that are essential for the normal function of the human nervous system.[31-33] Numerous factors, which include environment, lifestyle and genetics, result in an increased uptake of intracellular $Ca^{2+}$ that activates the protease calpain, truncating p35 into the fragments p10 and p25. The latter is a pathological activator of Cdk5, leading to its hyperactivation. Multiple cascading effects within the cell cycle can be traced back to the hyperactivation of Cdk5 resulting in the formation of β-amyloid plaques and intracellular neurofibrillary tangles – the well-known indicators of neurodegenerative diseases such as Alzheimer's disease (AD).[34-40]

Therapeutic approaches targeting Cdk5-related pathologies have focused on inhibitors such as aminothizole and roscovitine, which bind to the ATP-docking pocket and prevent Cdk5-mediated hyperphosphorylation.[41-44] However, targeting the ATP-binding pocket also causes non-specific interactions with other ATP-mediated cellular reactions, often causing serious side-effects. This has led to the pursuit of alternative approaches, such as the use of cholinesterase inhibitors[45] or antioxidants,[46,47] which have thus far not resulted in safe therapeutics.

In past work, we have shown a novel approach to inhibit the Cdk5-p25 pathology,[37] for example using the 24 amino acid, p5, obtained through the repeated truncation of p35. Importantly, these polypeptides act as selective inhibitors of Cdk5 pathological hyperactivity in both *in vivo* and *in vitro* experiments.[37,48,49] A variant derived from p5, TFP5, which was designed to cross the blood-brain barrier [50] showed a drastic decrease in pathology by allowing the rescue of cortical neurons in transgenic 5XFAD AD model mice *in vivo*.[40,50]

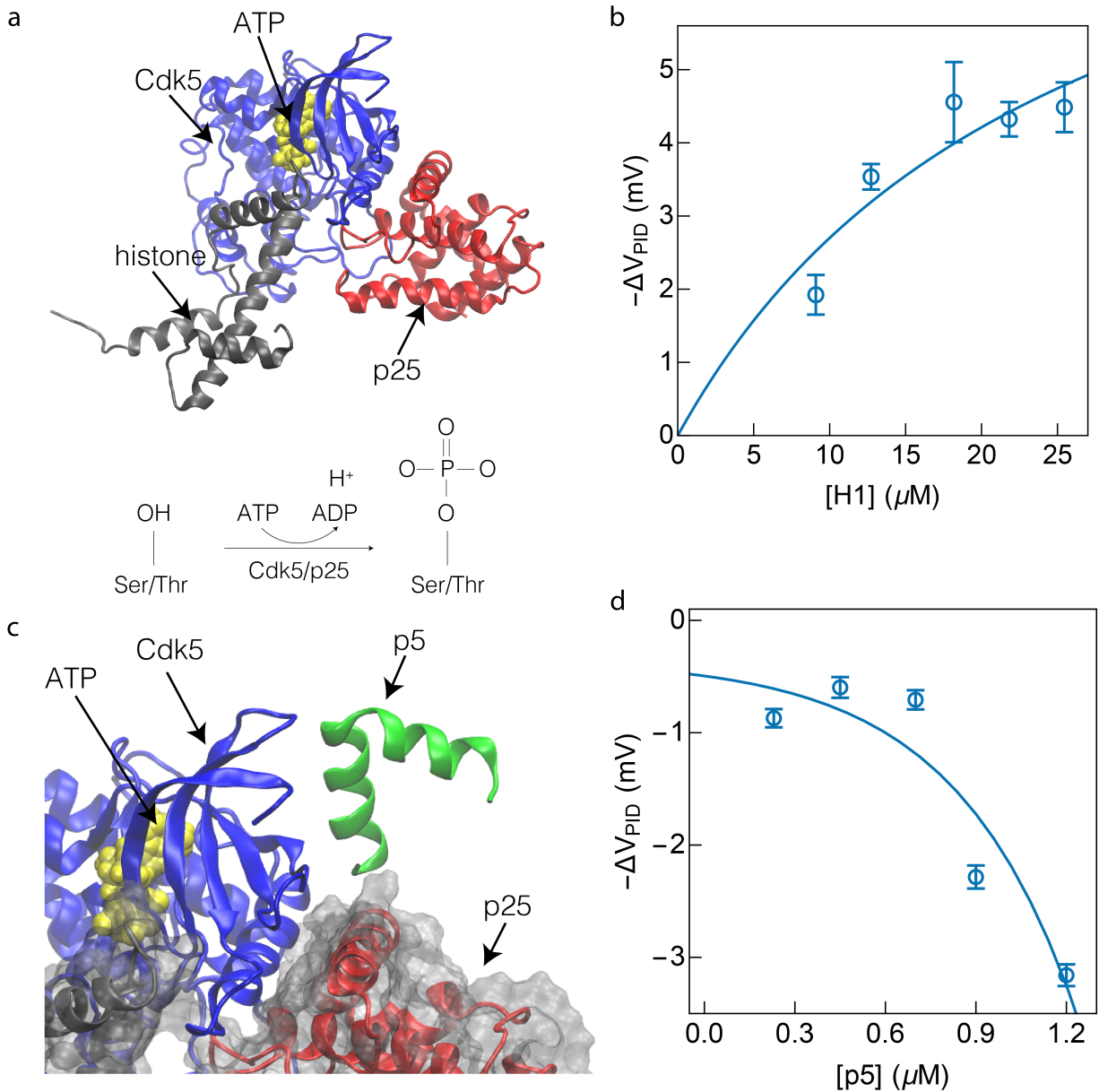

Figure 7: Measurements of the activity of the proline directed kinase, Cdk5 and the effect of the custom designed therapeutic polypeptide, p5, on modulating its activity. (a) (*top*) The molecular structure of the pathological Cdk5/p25 complex when phosphorylating a substrate protein, histone H1, during adenosine triphosphate (ATP) hydrolysis. (*bottom*) The reaction scheme of Cdk5-mediated phosphorylation of serine or threonine residues in histone. Upon hydrolysis of ATP, a single proton is released causing a slight acidification of the surrounding medium. (b) The change in the measured gate voltage ($\Delta V_{PID}$) of a n-channel silicon field-effect transistor (nFET) upon Cdk5-mediated phosphorylation as a function of the substrate protein, histone H1, concentration ([H1]). The solid line shows the fit of a Langmuir adsorption model to the data, which returned an activity coefficient, $k_a=(8.2\pm1.3)$ μM (c) Molecular representation of p5 interactions with the Cdk5/p25 complex that result in a decrease in its activity. (d) $\Delta V_{PID}$ as function of p5 concentration ([p5]) shows decreasing Cdk5 activity. The concentration of Cdk5/p25 and histone H1 were held a fixed value for each measurement in the plot. The solid line plots an exponential function to illustrate the trend in the data. The error bars in (b) and (d) represent the expanded uncertainty ($k=2$) in $\Delta V_{PID}$.

In ongoing work, we have used computer simulations to determine the molecular basis of p5-based inhibition mechanisms – a critical step towards developing safe therapeutics for the regulation of Cdk5/p25 hyperactivity.[51] This in turn could also lead to new molecules that are more selective and therefore safer. The FET-based measurements developed here can play a central role in this development cycle by enabling the rapid testing of candidate molecules. As a first step towards this goal, we demonstrate the ability of nFETs to measure the activity of Cdk5/p25 and the effect of p5 on re-regulation of this enzyme under physiological conditions.

**Enzymatic Activity of the Pathological Cdk5-P25 Complex.** Fig. 7a shows a molecular representation (protein data bank structure: 1UNL[52]) of the Cdk5-mediated phosphorylation (*top*) and the phosphorylation reaction scheme (*bottom*). In the presence of an activator protein (e.g., the pathological p25), Cdk5 catalyzes the transfer of a single phosphate group from ATP to a serine or threonine residue in a substrate protein (e.g., histone H1). The reaction also releases a single proton, thereby causing the surrounding medium to become slightly more acidic and decreasing its pH. When using the nFETs, the change in pH resulted in a change in $V_{PID}$ relative to a control sample with no enzyme, for concentrations of histone ([H1]) ranging from 9 µM to 25 µM as shown in Fig. 7b. All measurements were performed at a physiological concentration of Cdk5/p25 of 18.5 nM. As expected, the change in $V_{PID}$ increased monotonically with increasing histone concentration. A simple model of the form $\gamma \frac{[H1]}{k_a+[H1]}$, where $k_a$ is the activity and $\gamma$ is a scaling constant, was fit to the data in Fig. 7b. The value of $k_a$ was then estimated to be (8.2±1.3) µM when measured with the nFETs. This value was consistent with $k_a$=(9.1±0.9) µM for measurements performed using dg2DFET-based sensors (Fig. S1) under identical solution conditions, and also in agreement with our previous measurements using a $\gamma$-$^{32}$P-ATP assay ($k_a$=12.1±2.3 µM) as seen from Fig. S2.[8] In each case, the error bars of the estimated quantity represent the standard error of the measurement.

We leveraged the high time-resolution of our technique to measure the kinetics of the Cdk5/p25 enzymatic reaction for histone concentrations of 9.1 µM, 12.7 µM and 18.2 µM as seen from Fig. S3. In each case, the reaction was initiated upon the addition of ATP. After ≈1 minute, we observed a distinct change in the voltage signal that was indicative of histone phosphorylation. A control sample with no Cdk5/p25 showed no change in signal upon the addition of ATP. A nonlinear regression of the form $\beta(1 - e^{-k_1 t})$, where $\beta$ is a scaling constant, was used to estimate the enzyme rate constant $k_1$ to be (0.35±0.1) min$^{-1}$, consistent with literature values.[53]

**Inhibition of Cdk5/p25 activity with p5.** Upon the addition of the 24 amino acid polypeptide p5, we observed a strong reduction in Cdk5/p25 activity. A plausible structure of the complex formed between p5 and Cdk5/p25, derived from our previous computer simulations,[51] is shown in Fig. 7c. However, the molecular and regulatory mechanism of p5 action on Cdk5/p25 activity is still under investigation. As in Fig. 7b, the measurements were performed with a Cdk5/p25 concentration of 18.5 nM and a histone concentration of 25.4 µM. Fig. 7d shows the change in $V_{PID}$ when the p5 concentration ([p5]) was increased from 0.25 µM to 1.2 µM. From the figure, we can clearly see the effect of p5-based inhibition. Furthermore, the measurements agree with previous results of the interaction of p5 with the Cdk5/p25 complex obtained using a $\gamma$-$^{32}$P-ATP assay.[37] In particular, the sharp decrease in Cdk5/p25 activity past [p5]=0.7 µM. This decrease is indicative of a specific threshold for p5 inhibition and will be studied further in future work.

**Conclusions**

We show that the operation of commercially sourced nFETs can be optimized to achieve a pH resolution of $(7.2\pm0.3)\times10^{-3}$, ≈3-fold better than traditional ISFETs, and on par with solid-state dg2DFETs with an intrinsic gain $\alpha$=8. Furthermore, the improved performance is attained when the devices are operated in a remote configuration, with the pH sensing element located off-chip and connected electrically to the FET. This remote sensing approach is in contrast to conventional ISFETs which use integrated pH sensing membranes that form the gate dielectric. Our design greatly increases the versatility of nFETs in sensing applications, allowing them to be rapidly and easily interfaced with different biochemical sensors. It also allows the easy integration of other transduction elements into the biosensing setup that can be customized to meet the sensitivity, resolution and other performance needs of individual applications. Finally, the techniques outlined here can also be used to benefit applications that require integrated sensing, for example by improving the performance of traditional ISFETs.

The improved performance of the nFETs, operated in a remote configuration, was shown to be adequate to measure the activity of a pathological form of the proline directed kinase, Cdk5, which has been implicated to cause numerous neurodegenerative conditions including Alzheimer's disease. By using nFET-based measurements, we confirmed the effectiveness of a custom polypeptide, p5, in re-regulating Cdk5 function. Together the measurements demonstrate the feasibility of performing sensitive bioanalytical measurements with commercially available FETs, thereby drastically lowering the barrier to the adoption of this sensing technology.

**Conflicts of Interest**
The authors declare no conflicts of interest.

**Acknowledgements**

S.T.L. acknowledges support by the National Institute of Standards and Technology (NIST) grant 70NANB16H170. J.B.K. and N.B.G. acknowledge support by NIST grant 70NAHB15H023. A.C. acknowledges support by the NIST grant 70NANB17H259. Research performed in part at the NIST Center for Nanoscale Science and Technology nanofabrication facility.

**Footnotes**

†Certain commercial equipment, instruments, or materials are identified in this paper in order to specify the experimental procedure adequately. Such identifications are not intended to imply recommendation or endorsement by the National Institute of Standards and Technology, nor is it intended to imply that the materials or equipment identified are necessarily the best available for the purpose.